\documentclass[oneside,reqno,12pt]{amsart}
\usepackage{graphicx, amssymb, color}
\usepackage{amsmath}
\usepackage{graphicx}
\usepackage{enumerate}

\newtheorem{thm}{Theorem}[section]

\newtheorem{lem}[thm]{Lemma}
\newtheorem{prop}[thm]{Proposition}
\newtheorem{defn}[thm]{Definition}

\theoremstyle{remark}
\newtheorem{rem}[thm]{Remark}
\newtheorem{exa}[thm]{Example}

\newcommand{\prob}{\mathbb{P}}
\newcommand{\expec}{\mathbb{E}}
\newcommand{\CM}{\text{CM}}
\newcommand{\CR}{\text{CR}}
\newcommand{\EQ}{\text{EQ}}
\newcommand{\FX}{\text{FX}}
\newcommand{\IR}{\text{IR}}
\newcommand{\RF}{\text{RF}}
\newcommand{\LH}{\text{LH}}
\newcommand{\ES}{\text{ES}}

\newcommand{\FC}{\text{F,C}}
\newcommand{\RC}{\text{R,C}}
\newcommand{\RS}{\text{R,S}}
\newcommand{\IMCC}{\text{IMCC}}
\newcommand{\VaR}{\text{VaR}}
\newcommand{\SE}{\text{SE}}
\newcommand{\CAS}{\text{CAS}}

\begin{document}
\title[Capital allocation under the FRTB]{Capital allocation under \\the Fundamental Review of Trading Book}

\author[]{Luting Li}
\address[Luting Li]{Market Risk Analytics, 
Citibank N.A., London Branch,
London, UK; 
Department of Statistics,
London School of Economics and Political Science,
London, UK, Email: l.li27@lse.ac.uk}

\author[]{Hao Xing}
\address[Hao Xing]{Department of Statistics, London School of Economics and Political Science,
London, UK; Department of Finance,
Questrom School of Business, Boston University, Boston, USA, Email: h.xing@lse.ac.uk}

\begin{abstract}
Facing the FRTB, banks need to allocate their capital to each business units or risk positions to evaluate the capital efficiency of their strategies. This paper proposes two computationally efficient allocation methods which are weighted according to liquidity horizon. Both methods provide more stable and less negative allocations under the FRTB than under the current regulatory framework.
\smallskip

\end{abstract}

\date{\today}

\maketitle

\section{Introduction}

The Fundamental Review of Trading Book (FRTB) \cite{BCBS2016, BCBS2018} is a revised global risk management framework which aims to address shortcomings of the Basel II and its current amendments \cite{Basel25}. The FRTB sets out revised standards for minimum capital requirements for market risk and shifts from Value-at-Risk (VaR) to an Expected Shortfall (ES) measure.

In the new Internal Model Approach (IMA), tail risk and liquidity risk are considered and the capital-reducing effects of hedging are constrained. As a result, bank's global capital charge is facing significant changes.\footnote{In the industry Quantitative Impact Study (QIS-2014) \cite{QIS2014}, 44 banks report an average of 54\% increase of capital charge under the new IMA. In the QIS-2016 \cite{QIS2017}, 89 banks (including 71 Group1/G-SIBs banks and 18 Group 2 banks) report weighted average overall increases under the FRTB (IMA and SBA) by  51.7\% (51.4\% for Group 1/G-SIBs and 106.0\% for Group 2 banks) in market risk MRC.} It therefore becomes increasingly important for banks to re-evaluate the capital efficiency of their business structure. The first step of re-evaluation is to allocate firm-wide capital to each business unit or even each risk position. On the other hand, calculating the ES-based FRTB capital charge is computationally more demanding than calculating VaR under the current practice. Thus, in order to meet various risk management needs, new allocation methodology should be developed in a computationally efficient way.



We propose in this paper two allocation methods for the capital charge under the FRTB IMA. We focus on the risk factor and liquidity horizon bucketing, the liquidity horizon adjustment, and the stress period scaling, which are three main features of the FRTB. We highlight implications of these three features on capital allocation of modellable risk factor capital charge.\footnote{Capital charge under the FRTB IMA also incorporates the capital requirement for the non-modellable risk factors (NMRF) and the default risk charge (DRC). The NMRF capital charge is calculated as $\sqrt{\sum_{i=1}^L \text{ISES}^2_{NM, i}} + \sum_{j=1}^K \text{SES}_{NM,j}$ in \cite[Paragraph 190]{BCBS2016}. (An additional term related to the equity idiosyncratic risk is proposed in \cite{BCBS2018}. This additional term is similar to the first term in the previous formula.) The first term in the previous formula corresponds to NMRFs with $0$ correlation. Its allocation can be derived similarly as Lemmas \ref{lem:Euler} and \ref{lem:CAS} later. Allocation to the $j$-th NMRF from the second term in the previous formula can be  $\text{SES}_{NM,j}$ itself. Because \cite{BCBS2016} proposes to evaluate the DRC using a VaR framework, its allocation can be obtained using the existing Euler allocation.}  
Both allocation methods consist of two stages. In the first stage, the FRTB capital charge is allocated to different bucket of liquidity horizons and risk factors. Then, in the second stage, allocations in different buckets are decomposed, realigned, and aggregated again.

In the first allocation method, we examine
 the Euler allocation principle under the FRTB framework. The Euler allocation principle has been studied extensively. Tasche \cite{tasche1999risk} proves that the Euler allocation provides signal to optimise firm's portfolio return on risk-adjusted capital. Denault \cite{denault2001coherent} provides axiomatic characterisations of the Euler allocation. When the Euler allocation principle is applied under the FRTB framework, we show that the resulting allocation to each risk factor and liquidity horizon bucket is a scaled version of the standard Euler allocation. This scaling factor depends on the stand-alone ES of this bucket and the total FRTB ES of the same risk factor category. Our second allocation method is motivated by the constrained Aumann-Shapley allocation by Li et al. \cite{li2016organising}. Applying the Aumann-Shapley allocation to each risk factor category, we reduce the resulting allocation to another scaled version of the standard Euler allocation, where the scaling factor depends on the stand-alone ES of this bucket and its induced increment of FRTB ES. These two allocation methods are further extended, where the impact of additional risk positions on the stress period scaling factor is incorporated. Reducing the new allocation methods to the standard Euler allocation ensures computational efficiency. The same scenario extraction method can be used to compute the standard Euler allocation, without any revaluation of capital charges.

We illustrated our allocation methods via three groups of simulation analysis. Our analysis shows that risk factors with longer liquidity horizons are allocated with a larger proportion of the total FRTB capital charge. Secondly, negative allocations, resulting from hedging positions, in the Euler allocation of the standard ES are largely reduced or even reversed. Hedging between the different risk factor and liquidity horizon buckets rarely leads to negative allocations under the FRTB. Meanwhile hedging positions within the same bucket could still lead to negative allocations. However, magnitude of negative allocation to the same hedging position is much less in the FRTB than in the framework where the standard ES is evaluated on unconstrained P\&L.  Moreover, both allocation methods under the FRTB produce less dispersive allocations across different buckets than the Euler allocation of the standard ES. Therefore, both methods produce more stable allocations than the standard Euler allocation of the ES. Finally, our third simulation analysis demonstrates that allocation under the FRTB is sensitive to the choice of the reduced set of risk factors.

The rest of the paper is organised as follows: Section 2 introduces the expected shortfall measure under the FRTB  and investigates its homogeneity and sub-additivity properties. Allocation methods and their extensions are introduced in Section 3, followed by the simulation analysis in Section 4.

\section{FRTB expected shortfall}
\subsection{Risk factor and liquidity horizon bucketing}
Under the FRTB IMA framework, the P\&L of a risk position is attributed to risk factors (RF) of five different categories
\[
 \{\RF_i: 1\leq i\leq 5\}= \{\CM, \CR, \EQ, \FX, \IR\}.
\]
Each risk factor in the each category is assigned with a liquidity horizon (LH) with lengths
\[
\{\LH_j: 1\leq j\leq 5\} = \{10, 20, 40, 60, 120\}.
\]
Directly observable and frequently updated prices have shorter liquidity horizons. Risk factors associated to illiquid products and quantities which are calculated from direct observations typically have longer liquidity horizons.  A table of liquidity horizons of various risk factors is presented in \cite[Paragraph 181 (k)]{BCBS2016}.

We call \emph{negative} of the P\&L of a risk position the \emph{loss} of this risk position. The sign convention that positive value indicates the magnitude of loss will be employed throughout this paper. For a risk position $n$, $1\le n\le N$, we denote the \emph{constrained} 10-day loss of this position by a $5\times 5$ matrix $\tilde{X}_n = \{\tilde{X}_n(i,j)\}_{1\leq i, j \leq 5}$, where $\tilde{X}_n(i,j)$ is a random variable representing potential loss attributed to risk factors in $\RF_i$ with the liquidity horizon $\LH_j$. Netting among all RF and LH buckets, the 10-day loss of this position would be $\sum_{i,j}\tilde{X}_n(i,j)$. 

Now define the \emph{liquidity horizon adjusted} loss  as
\begin{equation}\label{def:Xn}
 X_n(i,j) = \sqrt{\frac{\LH_j- \LH_{j-1}}{10}}\sum_{k=j}^5 \tilde{X}_n(i,k), \quad 1\leq i, j\leq 5,
\end{equation}
where $\LH_0=0$. Considering the sum of losses attributed to all risk factors in the category $\RF_i$ with liquidity horizons at least as long as $\LH_j$, and scaling the sum by the factor $\sqrt{\frac{\LH_j -\LH_{j-1}}{10}}$, we obtain $X_n(i,j)$. We record the liquidity horizon adjusted bucketing of the risk position $n$ by a $5\times 5$ matrix $X_n = \{X_n(i,j)\}_{1\leq i, j\leq 5}$. We call the matrix $X_n$ as the \emph{risk profile} of the position $n$. Summing up all $\{X_n\}_{1\leq n\leq N}$, component-wise, we get the net risk profile of the portfolio 
\begin{equation}\label{def:X}
 X = \sum_n X_n,
\end{equation}
each component of $X$ is a random variable representing the net portfolio loss attributed to the bucket $(i,j)$.

The FRTB ES for the portfolio loss attributed to $\RF_i$ is defined in \cite[Paragraph 181 (c)]{BCBS2016} as
\begin{equation}\label{ES_i}
 \ES(X(i)) = \sqrt{\sum_{j=1}^5 \ES(X(i,j))^2},
\end{equation}
where each $\ES(X(i,j))$ is the expected shortfall of $X(i,j)$ calculated at the $97.5\%$ quantile.
\begin{exa}
 Consider a risk position whose loss is attributed only to  $\RF_i$  on $\LH_5=120$. Then $\tilde{X}(i,j) =0$ for any $j=1, \dots, 4$. Assume that the $10$ days loss $\tilde{X}(i,5)$ is normally distributed with zero mean and standard deviation $\sigma$. Then the loss over $120$ days is normally distributed with zero mean and standard deviation $\sqrt{120/10} \,\sigma$, hence its expected shortfall is $\sqrt{120/10} \,\sigma \ES(N(0,1))$, where $\ES(N(0,1))$ is the expected shortfall at the $97.5\%$ quantile of the standard normal distribution.  On the other hand, if we calculate  expected shortfall of the $120$ days loss via \eqref{ES_i}, we obtain the same expression. Indeed,  note that $X(i,j) = \sqrt{\frac{\LH_j - \LH_{j-1}}{10}} \tilde{X}(i,5)$, for $1\leq j\leq 5$. Then
 \[
  \ES(X(i)) = \sqrt{\sum_{j=1}^5 \frac{\LH_j - \LH_{j-1}}{10} \ES(\tilde{X}(i,5))^2} = \sqrt{\frac{120}{10}} \ES(N(0,\sigma)) = \sqrt{\frac{120}{10}} \, \sigma \ES(N(0,1)).
 \]
\end{exa}
\begin{rem}\label{rem:floor}
It is not required in \cite[Paragraph 181]{BCBS2016} to floor each $\ES(i,j)$ at zero. This means that negative $X(i,j)$ would lead to positive contribution in the risk measure $\ES(X(i))$.  Therefore, we suggest to floor each $\ES(X(i,j))$ at zero, and introduce 
\begin{equation}\label{ES+_i}
 \ES^+(X(i)) = \sqrt{\sum_{j=1}^5 \max\{\ES(X(i,j)), 0\}^2}.
\end{equation}
Our allocation methods introduced later can be applied to both $\ES(X(i))$ and $\ES^+(X(i))$.\end{rem}

\subsection{Stress period scaling and capital charge}
Besides the liquidity horizen adjustment, the FRTB introduces a scaling factor based on stress periods. For each risk factor category, calculate $\ES(X(i))$ in \eqref{ES_i} based on the current (most recent) $12$-month observation period with a full set of risk factors which are relevant to the risk position, and denote this risk measure as $\ES^{\FC}(X(i))$. Then identify a reduced set of risk factors, calculate its associated $\ES(X(i))$ over the same period, and denote it as $\ES^{\RC}(X(i))$. It is required that the reduced set of risk factors is large enough so that $\ES^{\RC}(X(i))$ is at least $75\%$ of $\ES^{\FC}(X(i))$. Subsequently, identify  a $12$-month stress period in which the portfolio experiences the largest loss,  calculate $\ES(X(i))$ with the reduced set of risk factors but use the observations from the stress period, and denote this risk measure as $\ES^{\RS}(X(i))$. FRTB IMA introduces the following expected shortfall capital charge  (see \cite[Paragraph 181 (d)]{BCBS2016}):
\begin{equation}\label{ES-IMA}
 \IMCC(X(i)) =\frac{\ES^{\RS}(X(i))}{\ES^{\RC}(X(i))} \ES^{\FC}(X(i)), \quad 1\leq i\leq 5.
\end{equation}

To consider the \emph{unconstrained} portfolio, we define 
\[
 X_n(6,j) = \sum_{i=1}^5 X_n(i,j), \quad 1\leq j\leq 5,
\]
which represents the net loss attributed to all risk factors from different categories but with the same liquidity horizon. We add $X_n(6, \cdot)$ as the $6$-th row in the risk profile and name the new $6\times 5$ matrix $X_n$ the \emph{extended} risk profile for the position $n$. Extending the risk profile of a portfolio similarly, we calculate $\IMCC(X(6))$ as \eqref{ES_i} and \eqref{ES-IMA} with $i=6$.

Now we are ready to introduce the capital charge for modellable risk factors under the FRTB IMA (see \cite[Paragraph 189]{BCBS2016}).
\begin{defn}\label{def:IMCC}
 The aggregate capital charge for modellable risk factors is a weighted sum of the constrained and unconstrained expected shortfall charges:
 \begin{equation}\label{IMCC}
  \IMCC(X) = \rho \,\IMCC(X(6)) + (1-\rho) \sum_{i=1}^5 \IMCC(X(i)),
 \end{equation}
 where the relative weight $\rho$ is set to be $0.5$.
\end{defn}

\subsection{Properties of \IMCC}\label{sec:IMCC-pro}
\begin{lem}\label{lem:IMCC-pro}
For any constant $a\geq 0$ and risk profiles $X$ and $Y$, the following statements hold:
\begin{enumerate}
 \item[(i)] (Positive homogeneity) \IMCC(aX) = a\,\IMCC(X).
 \item[(ii)] (Sub-additivity for \ES) For $i=1,\dots, 6$, if $\ES((X+Y)(i,j))\geq 0$ for any $j$, then
 \begin{equation}\label{ES-subadd}
  \ES((X+Y)(i)) \leq \ES(X(i)) + \ES(Y(i)).
 \end{equation}
 \item[(iii)] (Sub-additivity for \IMCC) For any $i=1, \dots, 6$, if
 \begin{equation}\label{cond1}
  \frac{\ES^{\RS}((X+Y)(i))}{\ES^{\RC}((X+Y)(i))} \leq \min\Big\{ \frac{\ES^{\RS}(X(i))}{\ES^{\RC}(X(i))},  \frac{\ES^{\RS}(Y(i))}{\ES^{\RC}(Y(i))}\Big\},
 \end{equation}
 and $\ES^{\FC}((X+Y)(i,j))\geq 0$ for any $j$, then
 \begin{equation}\label{IMCC-subadd}
  \IMCC((X+Y)(i)) \leq \IMCC(X(i)) + \IMCC(Y(i)).
 \end{equation}
\end{enumerate}
\end{lem}

Items (ii) and (iii) in the previous lemma present the sub-additivity property for the \ES\, and \IMCC\, capital charges under conditions \eqref{ES-subadd} and \eqref{cond1}. Without these conditions, the following examples show that the sub-additivity property may not hold.
\begin{exa}\label{exa-15}
 Consider two risk positions whose losses concentrate on $\RF_i$ and $\LH_j$. $X(i,j)$ has a Bernoulli distribution with $\prob(X(i,j) =-1) = \prob(X(i,j)=0)=0.5$, and $Y(i,j) = -1 - X(i,j)$. Hence $\prob((X+Y)(i,j) =-1) =1$. Then $\ES(X(i)) = \ES(Y(i)) =0$, but
 \[
  \ES((X+Y)(i)) = \big| \ES((X+Y)(i,j))\big| = |-1| =1 > \ES(X(i)) + \ES(Y(i)).
 \]
 However, if the expected shortfall is floored at zero as in Remark \ref{rem:floor}, then the sub-additivity property for \ES\, and \IMCC\, holds without the positivity assumption $\ES((X+Y)(i,j))\geq 0$ for all $j$.
\end{exa}

\begin{exa}
We consider two risk positions whose losses concentrate on $\RF_i$ and $\LH_j$. Assume that $X(i,j)$ and $Y(i,j)$ are i.i.d. standard normal, moreover,  the losses attributed to reduced sets account for $75\%$ and $100\%$, respectively, of the standard deviations of  the losses on full sets. Hence
\[
\ES^{\RC}(X(i))=0.75\ES^{\FC}(X(i)),\ \ES^{\RC}(Y(i))=\ES^{\FC}(Y(i)).
\]
Under stress scenarios, we assume that $X(i,j)$ and $Y(i,j)$ have independent normal distributions, but their standard deviations are scaled up by $1.2$ and $9$, respectively, of their values under current period. Then
\[
\min\Big\{ \frac{\ES^{\RS}(X(i))}{\ES^{\RC}(X(i))},  \frac{\ES^{\RS}(Y(i))}{\ES^{\RC}(Y(i))}\Big\}=\min\Big\{ 1.2,  9\Big\}=1.2.
\]
For the aggregated portfolio, the standard deviation of $X(i,j)+ Y(i,j)$ attributed to the full set is $\sqrt{2}$, and $\sqrt{0.75^2 +1} = 1.25$ to the reduced set. Under the  stress scenarios, the standard deviation of $X(i,j)+ Y(i,j)$ attributed to the reduced set becomes $\sqrt{(0.75\times 1.2)^2+9^2}\approx 9.04$. Hence
\[
 \frac{\ES^{\RS}((X+Y)(i))}{\ES^{\RC}((X+Y)(i))}=\frac{9.04}{1.25}=7.23>1.2.
\]
Therefore, the condition \eqref{cond1} is violated. Now we have
\begin{align*}
 \IMCC((X+Y)(i))&= \frac{\ES^{\RS}((X+Y)(i))}{\ES^{\RC}((X+Y)(i))}\ES^{\FC}((X+Y)(i))\\
 &=7.23\times \sqrt{2}ES(N(0,1)).
\end{align*}
On the other hand, comparing with the sum of two IMCCs that
\begin{align*}
 \IMCC(X(i))+ \IMCC(Y(i))&= \frac{\ES^{\RS}(X(i))}{\ES^{\RC}(X(i))}\ES^{\FC}(X(i))+ \frac{\ES^{\RS}(Y(i))}{\ES^{\RC}(Y(i))}\ES^{\FC}(Y(i))\\
 &=(1.2+9)ES(N(0,1)),
\end{align*}
we find
\begin{equation*}\label{violate-IMCC-subadd}
 7.23\times \sqrt{2}=10.22>10.20.
\end{equation*}
Hence \eqref{IMCC-subadd} fails.
\end{exa}

\section{Capital allocation}

We introduce in this section several methods to allocate the aggregated capital charge IMCC to different components of a portfolio. All allocation methods have two steps. 
Given a risk measure $RM$ and an extended portfolio risk profile $X$, the first step allocates capital to each bucket $X_n(i,j)$. Denote the allocation to $X_n(i,j)$ from the total capital $RM(X)$ by
\[
 RM (X_n(i,j) | X).
\]
Recall from \eqref{def:Xn} that $X_n(i,j)$ is aggregated from $\tilde{X}_n(i,k)$ with $k\geq j$. In the second step, we reverse the liquidity horizon adjustment to further allocate $RM(X_n(i,j) | X)$ to $\tilde{X}_n(i,k)$ and denote the resulting allocations by 
\[RM (\tilde{X}_n(i,k) | X_n(i,j)), \quad k\geq j.\]
Finally, we sum up all contributions from $X_n(i,j)$ with $j\leq k$ to obtain the allocation for $\tilde{X}_n(i,k)$:
\begin{equation}\label{sec-step}
 RM(\tilde{X}_n(i,k) | X) = \sum_{j=1}^k RM(X_n(i,k) | X_n(i,j)).
\end{equation}
In all methods, the second step is the same, we will focus on the first step in what follows.



\subsection{Euler allocation}
Euler allocation has been studied extensively; see \cite{litterman1996hot}, \cite{tasche1999risk}, \cite{denault2001coherent}, \cite{tasche2007capital}, and many others. We introduce in this section a computationally efficient scheme for Euler allocation of the IMCC.

For each $\RF_i$, we first allocate $\ES(X(i))$ in \eqref{ES_i} to each $X_n(i,j)$. Let us introduce some notation.
Let $v= (v_n)_{1\leq n\leq N}$ be a sequence of real numbers as weights.  Given a collection of risk profiles $\{X_n\}_{1\leq n\leq N}$, denote
\begin{equation}\label{Xvj}
 X^{v,j}(i)= \sum_n X^{v_n, j}_n(i),
\end{equation}
where the sum is computed component-wise and
\[
 X^{v_n, j}_n(i) = \big(X_n(i, 1), \cdots, X_n(i, j-1), v_n X_n(i, j), X_n(i, j+1), \cdots, X_n(i, 5)\big),
\]
i.e. the weight $v_n$ is put on $X_n(i,j)$ but unit weight is put on all other LHs.
For each $\RF_i$, we define the allocation to each $X_n(i,j)$ as follows.

\begin{defn}[Euler allocation of FRTB ES]
 For $1\leq n\leq N, 1\leq i \leq 6, 1\leq j\leq 5$, let
 \begin{equation}\label{Euler-der-0}
  \ES(X_n(i,j) \,|\, X(i)) := \frac{\partial}{\partial v_n} \ES(X^{v,j}(i)) \Big|_{v=1},
 \end{equation}
 where $\ES(X^{v,j}(i))$ is the FRTB ES of the row $X^{v,j}(i)$ in \eqref{Xvj}, and $v=1$ represents $v_n=1$ for all $n$. We call $\ES(X_n(i,j) \,|\, X(i))$ the Euler allocation of FRTB ES.
\end{defn}

The chain rule in differentiation yields the following representation.
\begin{lem}\label{lem:Euler}
For $1\leq n\leq N, 1\leq i \leq 6, 1\leq j\leq 5$,
\begin{equation}\label{Euler-der}
 \ES(X_n(i,j) \,|\, X(i)) = \frac{\ES(X(i,j))}{\ES(X(i))} \, \frac{\partial}{\partial v_n} \ES\big(X^v(i,j)\big) \Big|_{v=1},
\end{equation}
where $X^v(i,j) = \sum_n v_n X_n(i,j)$.
\end{lem}

Note that $\partial_{v_n} \ES(X^v(i,j)) \big|_{v=1}$ in \eqref{Euler-der} is the standard Euler allocation of $\ES(X(i,j))$. Then the Euler allocation under FRTB ES is the weighted version of the standard Euler allocation. The scaling factor $\frac{\ES(X(i,j))}{\ES(X(i))}$ reflects the ratio between the stand-alone ES of $X(i,j)$ and the FRTB ES of $X(i)$. This scaling factor is applied to all risk positions of the same liquidity horizon.

When the distribution of $X(i,j)$ satisfies certain regularity conditions (cf. \cite[Assumption (S)]{tasche1999risk}), then the standard Euler allocation can be calculated as a conditional expectation (cf. \cite{tasche1999risk}):
\begin{equation}\label{Euler-SE}
 \frac{\partial}{\partial v_n} \ES\big(X^v(i,j)\big) \Big|_{v=1} = \expec\big[X_n(i,j) \,|\, X(i,j) \geq \VaR(X(i,j))\big] =: \SE\big(X_n(i,j) \,|\, X(i,j)\big),
\end{equation}
where $\VaR(X(i,j))$ is the Value-at-Risk of $X(i,j)$ calculated at the $97.5\%$ quantile. The conditional expectation above can be calculated by the \emph{scenario-extraction} method and hence is denoted by $\SE\big(X_n(i,j) \,|\, X(i,j)\big)$.  Applying the scaled scenario-extraction method to \eqref{Euler-der} is also computationally efficient. Rather than calculating the element-wise derivative in \eqref{Euler-der-0} using a numeric differential scheme
\[
\frac{\partial}{\partial v_n(i,j)} \ES(X^v(i)) \Big|_{v=1} = \lim_{\epsilon\downarrow 0} \frac{1}{\epsilon} \Big(\ES\big(X(i) + \epsilon X_n(i,j)\big) - \ES(X(i))\Big),
\]
which typically requires revaluation on the bumps for each position, the scenario-extraction method calculates the conditional expectation by averaging $X_n(i,j)$ on scenarios when the portfolio loss $X(i,j)$ violates $\VaR(X(i,j))$.

After applying the Euler allocation to the FRTB ES under full set of risk factors, and scaling the allocations by the stress period scaling factor, we have the following allocation to the IMCC capital charge.

\begin{defn}[Euler allocation of IMCC]\label{def:Euler-IMCC}
 For $1\leq n\leq N, 1\leq i \leq 6, 1\leq j\leq 5$, let
 \begin{equation}\label{Euler-IMCC}
\IMCC^E(X_n(i,j) \, |\, X(i)) := 0.5 \,\frac{\ES^{\RS}(X(i))}{\ES^{\RC}(X(i))} \, \ES^{\FC} \big(X_n(i,j) \,|\, X(i)\big).
 \end{equation}
 We call $\IMCC^E(X_n(i,j) \, |\, X(i))$ the Euler allocation of IMCC. For the risk profile $X_n$ of the risk position $n$, we define its Euler allocation as 
 \[
  \IMCC^E(X_n \, |\, X) = \sum_{i,j} \IMCC^E \big(X_n(i,j) \, |\, X(i)\big).
 \]
\end{defn}

\begin{prop}\label{prop:Euler-IMCC}
 The Euler allocation of IMCC is a full allocation, i.e.,
 \[
 \sum_n \IMCC^E(X_n\,|\, X) = \sum_{n,i,j} \IMCC^E\big(X_n(i,j) \,|\, X(i)\big) = \IMCC(X).
 \]
\end{prop}

\begin{rem}\label{rem:Euler-floor}
 If the expected shortfall for $X^v(i,j)$ is floored at zero as in Remark \ref{rem:floor}, \eqref{Euler-der} can be replaced by
 \[
 \ES^+(X_n(i,j) \,|\, X(i)) =\left\{\begin{array}{ll} \frac{\ES^+(X(i,j))}{\ES^+(X(i))} \, \SE\big(X_n(i,j) \,|\, X(i,j)\big) & \text{ if } \ES(X(i,j))>0 \\ 0 & \text{ otherwise}\end{array}\right..
 \]
 The resulting Euler allocation of IMCC is still a full allocation, since  $\ES^+$ is still homogeneous of degree $1$.
\end{rem}

When a portfolio contains sub-portfolios which hedge each other, the standard Euler allocation under expected shortfall could produce negative allocations to some sub-portfolios. Because the FRTB ES discourages hedging across different risk factor classes and different liquidity horizons, negative allocations could be reduced or reversed under the FRTB. The following example illustrates this point.
\begin{exa}\label{exa:neg-all}
 Consider a portfolio with two risk positions whose risk profiles are denoted by $Y$ and $Z$, respectively. We assume that $Y$ concentrates on $\RF_i$ and $\LH_j$, and $Z$ concentrates on $\RF_k$ and $\LH_j$, with $1\leq i\neq k\leq 5$. Therefore, the matrix-valued random variables $Y$ and $Z$ concentrate on their $(i,j)$-th and $(k,j)$-th components $Y(i,j)$ and $Z(k,j)$, respectively. We consider a hypothetical situation that $Y(i,j)=-Z(k,j)$ and both of them follow standard normal distributions. Then the net loss of the portfolio $X$ is zero, and the standard Euler allocation of ES would be negative for either $Y(i,j)$ or $Z(k,j)$, say $\SE(Y(i,j)|X)<0$. 
 
However, under the FRTB framework, $X(i) = Y(i,j)$ and $X(k)=Z(k,j)$. Then 
\begin{align*}
\IMCC^E\big(Y(i,j) \, |\, X(i)\big) &= 0.5 \frac{\ES^{\RS}(Y(i,j))}{\ES^{\RC}(Y(i,j))} \ES^{\FC}(Y(i,j) | X(i)) \\
&= 0.5 \frac{\ES^{\RS}(X(i))}{\ES^{\RC}(X(i))} \ES^{\FC}(X(i))>0.
\end{align*}
In a more realistic situation, $Y$ and $Z$ are unlikely cancelling each other, but a sufficiently negative correlation introduces negative allocations in the standard Euler allocation. Under the FRTB framework, since they are associated to different risk factor classes, allocations to each constrained classes are always positive. These positive allocations would compensate potential negative allocations in the unconstrained classes. Therefore, $\IMCC^E(Y|X)$ could be less negative, or even positive.
\end{exa}

\subsection{Constrained Aumann-Shapley allocation}
The Shapley and Aumann-Shapley allocations were introduced in \cite{denault2001coherent}, where the results of \cite{shapley1953} and \cite{aumann1974values}  on coalitional games were applied to capital allocation problems. The concepts in those two allocations were combined in  \cite{li2016organising} to introduce the Constrained Aumann-Shapley allocation, where permutations of different risk positions are restricted to each business unit. In the FRTB IMA framework,  the risk factor bucketing rule produces a natural constraint on risk profile organisations. Therefore we constrain the Shapley-type permutations within each RF classes.

We introduce the following full permutation matrix:
$$
\mathcal{L}:=\left[
\begin{matrix}
10 & 20 & 40 & 60 & 120\\
10 & 20 & 40 & 120 & 60\\
\vdots & \vdots & \ddots & \vdots & \vdots\\
120 & 60 & 40 &20 & 10
\end{matrix}
\right]_{5! \times 5}.
$$
Each row of $\mathcal{L}$ records a permutation of liquidity horizons $\{10, 20, 40, 60, 120\}$. There are $5! =120$ permutations in total. For a given row $r$ and a liquidity horizon $\LH_j$, we denote $\mathcal{L}^{-1}(r, j)$ the column of $\mathcal{L}$ in which $\LH_j$ locates. For example, $\mathcal{L}^{-1}(2, 5)= 4$, or equivalently, $\mathcal{L}(2,4)=\LH_5=120$.

Given a risk profile $X_n$, a risk factor category $\RF_i$, a liquidity horizon $\LH_j$, and a permutation of liquidity horizons (say  $r$-th row in $\mathcal{L}$). We want to first allocate $\ES(X(i))$ to $X_n(i,j)$. We call this allocation as the \emph{Constrained Aumann-Shapley} (CAS) allocation of FRTB ES, and denote it as
\[
 \CAS(r, X_n(i,j)).
\]
To introduce the value of $\CAS(r, X_n(i,j))$, let $v=(v_n)_{1\leq n\leq N}$ be a sequence of real numbers representing weights and
\begin{equation}\label{Xvrj}
 X^{v, r, j}(i) = \sum_n X^{v, r, j}_n(i),
\end{equation}
where $X^{v, r, j}_n(i)$ is a row with the entry $X_n(i, \ell)$ at the $\ell$-th column if $\mathcal{L}^{-1}(i, \ell) < \mathcal{L}^{-1}(i, j)$ (i.e., $\LH_\ell$ appears before $\LH_j$ in the permutation $r$); the entry $v_n X_n(i,j)$ at the $j$-th column; and zero in all other columns. Taking the second row in matrix $\mathcal{L}$ as an example, for $j=5$ we have
\[
X^{v, 2, 5}_n(i)=  \big(X_n(i, 1),X_n(i, 2), X_n(i,3), 0, v_n X_n(i, 5)).
\]
Then define
\[
 \CAS(r, X_n(i,j)) := \int_0^1 \frac{\partial}{\partial v_n} \ES(X^{v,r,j}(i))\Big|_{ v=q} \,dq,
\]
where $v=q$ means $v_n=q$ for all $n$. Intuitively, $\partial_{v_n} \ES(X^{v, r,j}(i)) |_{v=q}$ is the marginal contribution, in the direction of $X_n(i,j)$, of the FRTB ES for the portfolio risk profile consisting $q X(i, j)$ and all $X(i, \ell)$ whose liquidity horizon $\LH_\ell$ appears before $\LH_j$ in the permutation $r$.

\begin{lem}\label{lem:CAS}
For $1\leq n\leq N$, $1\leq i\leq 6$, $1\leq j\leq 5$, and $1\leq r\leq 5!$,
\begin{equation}\label{CAS-der}
 \CAS(r, X_n(i,j)) = \eta(r, i, j)\, \frac{\partial}{\partial v_n} \ES\big(X^v(i,j)\big) \Big|_{v=1},
\end{equation}
where
\begin{equation}\label{eta}
 \eta(r, i, j) = \frac{\sqrt{\sum_{1\leq s\leq \mathcal{L}^{-1}(r, j)} \ES\big(X(i, \mathcal{L}(r,s))\big)^2} - \sqrt{\sum_{1\leq s < \mathcal{L}^{-1} (r, j)} \ES\big(X(i, \mathcal{L}(r, s))\big)^2}}{\ES\big(X(i,j)\big)}.
\end{equation}
When the distribution of $X(i,j)$ satisfies \cite[Assumption (S)]{tasche1999risk}, then the derivative on the right-hand side of \eqref{CAS-der} can be replaced by $\SE\big(X_n(i,j) \,|\, X(i,j)\big)$ in \eqref{Euler-SE}.
\end{lem}

Similar to the Euler allocation under FRTB ES, the Constrained Aumann-Shapley allocation is also a weighted version of the standard Euler allocation. The scaling factor $\eta(r,i,j)$ is the ratio between the $X(i,j)$ induced incremental FRTB ES in the permutation $r$ and the stand-alone ES of $X(i,j)$.

After averaging over all permutations, we introduce the following allocation to the IMCC capital charge.
\begin{defn}[CAS allocation of IMCC]\label{def:CAS-IMCC}
For $1\leq n\leq N$, $1\leq i\leq 6$, $1\leq j \leq 5$,
\[
\IMCC^C(X_n(i,j) \,|\, X(i)) := 0.5 \,\frac{\ES^{\RS}(X(i))}{\ES^{\RC}(X(i))} \, \frac{1}{5!}\sum_{r=1}^{5!} \CAS^{\FC}(r, X_n(i,j)),
\]
where $\CAS^{\FC}$ is the Constrained Aumann-Shapley allocation of FRTB $ES^{\FC}$.
We call $\IMCC^C(X_n(i,j) \,|\, X(i))$ the Constrainted Aumann-Shapley allocation of IMCC.
\end{defn}
\begin{prop}\label{prop:CAS-IMCC}
The CAS allocation of IMCC is a full allocation, i.e.
 \[
 \sum_n \IMCC^C(X_n\,|\, X) = \sum_{n,i,j} \IMCC^C\big(X_n(i,j) \,|\, X(i)\big) = \IMCC(X).
 \]
 \end{prop}
 If the expected shortfall for $X^v(i,j)$ is floored at zero as in Remark \ref{rem:floor}, the CAS allocation can be adjusted similarly to Remark \ref{rem:Euler-floor}. The adjusted CAS allocation is still a full allocation.

\begin{rem}
\label{rmk:additivity}
An important concept for capital allocation is the \emph{additivity} property. Consider a subportfolio $Y$ in $X$, where $Y$ is aggregated from risk profiles $\left\{Y_m\right\}_{1\le m \le M}$. We want to know whether the allocation to the portfolio $Y$ equals to the sum of allocations to all $\{Y_m\}$, i.e. whether
$\rho\left(Y|X\right)=\sum_{m}\rho\left(Y_m|X\right)$ is true.
The answer to this question is positive for both Euler and CAS allocations. This is due to the fact that both of them are scaled versions of the Euler allocation for the regular ES, which is additive itself.
\end{rem}


\subsection{The second step allocation}
\label{rmk:tildeXallocation}
After the first step of both allocation methods, capital is allocated to  each liquidity horizon adjusted loss $X_n(i,j)$. For the unconstrained part $i=6$, we consider $X_n(6,j)=\sum_{i=1}^5 X_n(i,j)$ and use the standard Euler allocation to allocate unconstrained allocation to each $X_n(i,j)$ and denote it by $\IMCC(X_n(i,j)\,|\, X(6))$.

Now for each $1\leq i\leq 6$, since $X_n(i,j)$ is aggregated from $10$ days loss $\tilde{X}_n(i,k)$ with $k\geq j$, it seems natural to extract capital associated to each $\tilde{X}(i,k)$ from the capital allocated to $X(i,j)$. Recall from \eqref{def:Xn}. We can consider $X_n(i,j)$ as a portfolio of $\sqrt{\frac{\LH_j -\LH_{j-1}}{10} }\tilde{X}_n(i,k)$ with $k\geq j$. Hence we use the Euler method to allocate capital from $X_n(i,j)$ further down to each  $\sqrt{\frac{\LH_j -\LH_{j-1}}{10} }\tilde{X}_n(i,k)$. We denote the resulting allocations by
$$
\IMCC\left(\sqrt{\frac{\LH_{j}-\LH_{j-1}}{10}}\tilde{X}_n(i,k)\bigg| X_n(i,j)\right), \quad k\geq j.
$$
Now using the additivity property in Remark \ref{rmk:additivity}, we can sum all capital from $X_n(i,j)$ with $j\leq k$ to get the contribution of $\tilde{X}_n(i,k)$ as
\begin{equation}\label{IMCC-tX}
\IMCC\left(\tilde{X}_n(i,k)|X(i)\right)=\sum_{j\leq k}\IMCC\left(\sqrt{\frac{\LH_{j}-\LH_{j-1}}{10}}\tilde{X}_n(i,k)\bigg| X_n(i,j)\right).
\end{equation}
Combining constrained and unconstrained allocations, the allocation for $\tilde{X}_n(i,j)$, with $1\le n\le N$, $1\le i \le 5$ and $1\le j\le 5$,  is given by
\begin{equation}\label{IMCC-total-allocation}
\IMCC^{Total}\left(\tilde{X}_n(i,k)|X(i)\right):=\IMCC\left(\tilde{X}_n(i,k)|X(i)\right)+\IMCC\left(\tilde{X}_n(i,k)|X(6)\right).
\end{equation}

\subsection{Extensions}\label{sec:scal-adj}
In the previous two sections, Euler and CAS allocations of IMCC are applied to the FRTB ES for the Full Current set, and the stress scaling factor $\frac{\ES^{\RS}(X(i))}{\ES^{\RC}(X(i))}$ is treated as a constant for each $\RF_i$. In other words, the $X_n(i,j)$ induced risk contribution is considered for $\ES^{\FC}$, but \emph{not} for $\ES^{\RS}$ and $\ES^{\RC}$. In this section, we will consider the impact of $X_n(i,j)$ on the stress scaling factors. The second step of allocation is the same as in Section \ref{rmk:tildeXallocation}.
\begin{defn}[Euler allocation of IMCC with scaling adjustment]\label{den:defn3-11}
For $1\leq n\leq N$, $1\leq i\leq 6$, $1\leq j\leq 5$, let
\[
\IMCC^{\text{E,S}}\big(X_n(i,j) \, |\, X(i)\big) := 0.5 \frac{\partial}{\partial v_n} \Big[\frac{\ES^{\RS} \big(X^{v,j}(i)\big)}{\ES^{\RC} \big(X^{v,j}(i)\big)} \ES^{\FC}\big(X^{v,j}(i)\big)\Big] \Big|_{v=1}.
\]
\end{defn}
Taking differentiations to each expected shortfalls, we obtain
\begin{prop}\label{prop:stressallocationnewlyadded}
 For $1\leq n\leq N$, $1\leq i\leq 6$, $1\leq j\leq 5$,
\begin{equation}\label{IMCC-ES}
\begin{split}
\IMCC^{\text{E,S}}\big(X_n(i,j) \, |\, X(i)\big) = 0.5 \Big[&\frac{\ES^{\RS}(X(i))}{\ES^{\RC}(X(i))} \, \ES^{\FC} \big(X_n(i,j) \,|\, X(i)\big)\\
+ & \frac{\ES^{\FC}(X(i))}{\ES^{\RC}(X(i))} \, \ES^{\RS} \big(X_n(i,j) \,|\, X(i)\big) \\
- &\frac{\ES^{\RS}(X(i))\ES^{\FC}(X(i))}{\ES^{\RC}(X(i))^2} \, \ES^{\RC} \big(X_n(i,j) \,|\, X(i)\big)\Big].
\end{split}
\end{equation}
\end{prop}
The previous expression for $\IMCC^{\text{E,S}}$ motivates us to define the following CAS allocation with scaling adjustment.

\begin{defn}
For $1\leq n\leq N$, $1\leq i\leq 6$, $1\leq j\leq 5$, let
\begin{align*}
 \IMCC^{\text{C,S}}\big(X_n(i,j) \,|\, X(i)\big) := \frac{0.5}{5!} \sum_{r=1}^{5!} \Big[&\frac{\ES^{\RS}(X(i))}{\ES^{\RC}(X(i))} \, \CAS^{\FC}\big(r, X_n(i,j)\big) \\
+ &\frac{\ES^{\FC}(X(i))}{\ES^{\RC}(X(i))} \, \CAS^{\RS}\big(r, X_n(i,j)\big) \\
- &\frac{\ES^{\RS}(X(i))\ES^{\FC}(X(i))}{\ES^{\RC}(X(i))^2} \, \CAS^{\RC}\big(r, X_n(i,j)\big) \Big].
\end{align*}
\end{defn}
\begin{prop}\label{prop:Euler-CAS-ad}
 Both Euler and CAS allocations of IMCC with scaling adjustment are full allocations and satisfy the additivity property.
\end{prop}

\section{Simulation Analysis}
\subsection{Positive correlations}\label{subsec:sim1}
This simulation exercise illustrates the difference of allocations among different RFs and LHs. We assume that there is only one risk position, and all $\tilde{X}(i,j)$ have identical normal distributions with zero mean and 30\% annual volatility. We consider the following four scenarios of correlation structures:
\begin{enumerate}
 \item[(i)] Independence: all $\tilde{X}(i,j)$ are mutually independent;
 \item[(ii)] Uniform positive correlation: each pair of $\tilde{X}(i,j)$ and $\tilde{X}(k,l)$ have correlation $0.99$;
 \item[(iii)] Positive correlation among RFs and zero correlation among LHs: $\text{corr}(\tilde{X}(i,j), \tilde{X}(k,j))=0.99$ and $\text{corr}(\tilde{X}(i,j), \tilde{X}(i,k))=0$ for any $i\neq k$;
 \item[(iv)] Positive correlation among LHs and zero correlation among RFs: $\text{corr}(\tilde{X}(i,j), \tilde{X}(k,j)) =0$ and $\text{corr}(\tilde{X}(i,j), \tilde{X}(i,k))= 0.99$ for any $i\neq k$.
\end{enumerate}

This exercise assumes extreme correlations among different RFs and LHs to highlight their impact to FRTB allocations. When correlations are moderate, similar patterns appear but are less pronounced.

We simulate 250 times independent 10 day-loss. In each day, different correlation structures are specified as in above. The stress period scalings are assumed to be $1$ for all \RF s. First, we compare the IMCC and the $97.5\%$ ES of net loss distribution without distinguishing RFs and LHs in the following table. We call this 97.5\% ES \emph{regular} ES in what follows. Besides, the comparisons below should not be understood as the QIS-style exercise, as the RWA under the current Basel 2.5/3 practices are based on the VaR metric. The example here aims to show the capital impact from the FRTB IMA by only considering the RF and LH constraints, and the regular ES is considered for the benchmark purpose.

\begin{table}[h]
\centering
  \begin{tabular}{ c | c  c  }
    \hline
     Scenario& IMCC &  Regular ES\\ \hline
    (i) Independent & 12.48 & 3.28 \\ 
    (ii) Uniform Positive Corr   & 28.57 &  16.70 \\
    (iii) Positive-RF-Corr & 18.28 & 7.81 \\
    (iv) Positive-LH-Corr & 21.00 & 7.59 \\
    \hline
    \hline
  \end{tabular}
  \caption {FRTB IMCC v.s. Regular ES}\label{mytable_frtb_es}
\end{table}

We can see from Table \ref{mytable_frtb_es} that the IMCC values are between 1.7 and 3.8 times of the regular ES. Moreover, comparing scenarios (iii) and (iv), we see that positive correlations between different LHs increase IMCCs more than positive correlations between different RFs. This is due to the FRTB \LH\ scaling rule in Equation \eqref{def:Xn}.

\begin{figure}[h]
 \centering
    \includegraphics[height=0.73\textwidth,width=0.9\textwidth]{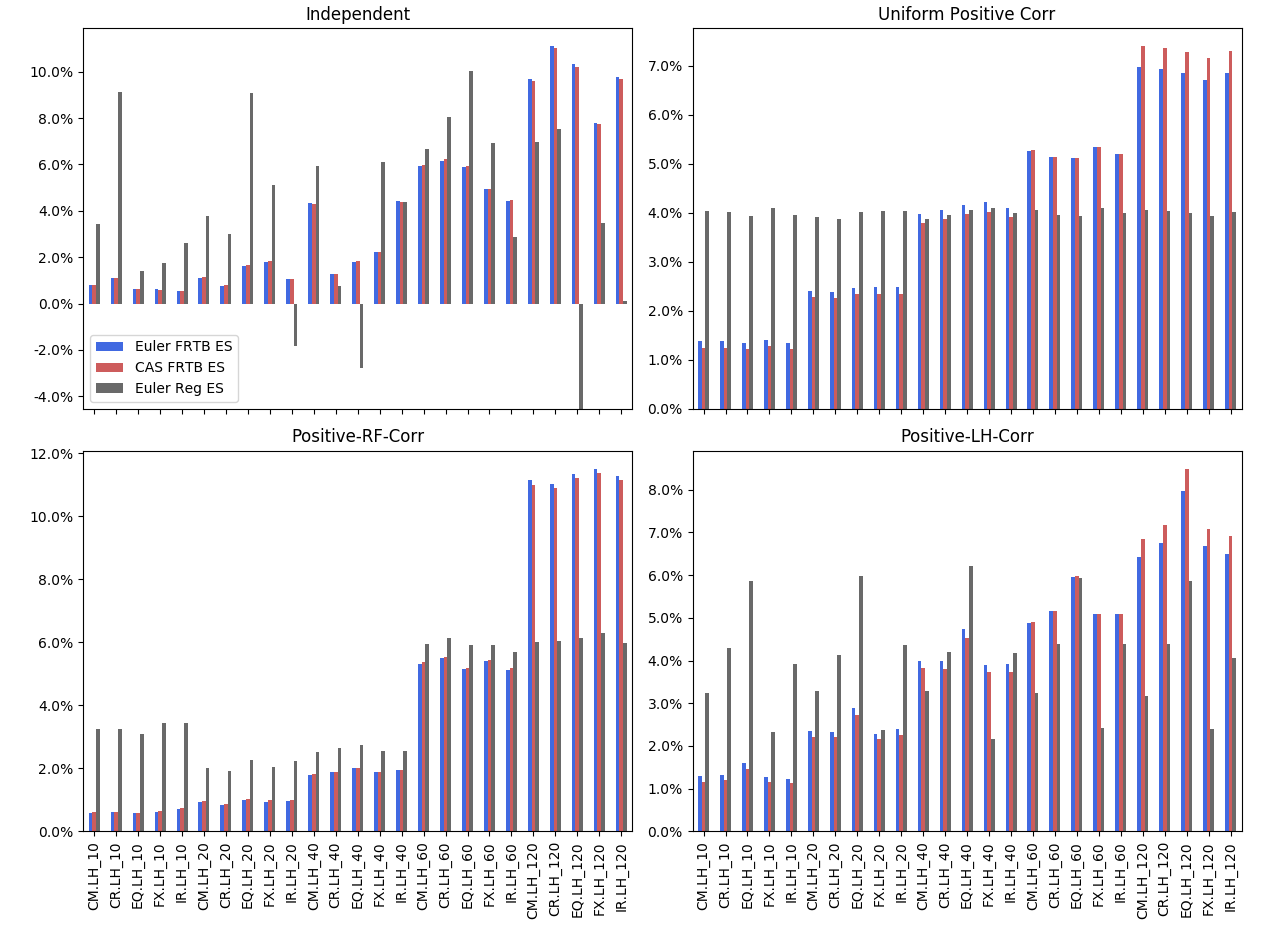}
\caption{Euler allocation of FRTB ES (Euler FRTB ES), CAS allocation of FRTB ES (CAS FRTB ES), and Euler allocation of Regular ES (Euler Reg ES). Upper-left panel: scenario (i); Upper-right panel: scenario (ii); Bottom-left panel: scenario (iii); Bottom-right panel: scenario (iv). Each panel presents the percentage of allocation to different $\tilde{X}(i,j)$. The total capital charges are reported in  Table \ref{mytable_frtb_es}.
}
\label{NonKernelFigs}
\end{figure}

Figure \ref{NonKernelFigs} illustrates  the Euler allocation of FRTB ES, the CAS allocation of FRTB ES, and the Euler allocation of regular ES. It reports allocations to different $\tilde{X}(i,j)$, after combining the constrained and unconstrained allocations (see Equation  \eqref{IMCC-total-allocation}).

Figure \ref{NonKernelFigs} shows that both FRTB allocation methods typically allocate more capital to risk factors with longer liquidity horizons. This feature is due to the facts that 1) longer liquidity horizon has bigger scalings (see Equation \eqref{def:Xn}); and 2) longer liquidity horizon has more allocation contributions from shorter liquidity horizon allocations (see Equation \eqref{IMCC-tX}). On the other hand, due to allocations from unconstrained part, when there is no strong positive correlation among risk factor categories, allocations to each liquidity horizon vary within the same risk factor category. However, the regular ES Euler allocation does not show a consistent pattern. This is because losses are aggregated without distinguishing different RFs and LHs.

\begin{figure}[h]
 \centering
    \includegraphics[height=0.73\textwidth,width=0.9\textwidth]{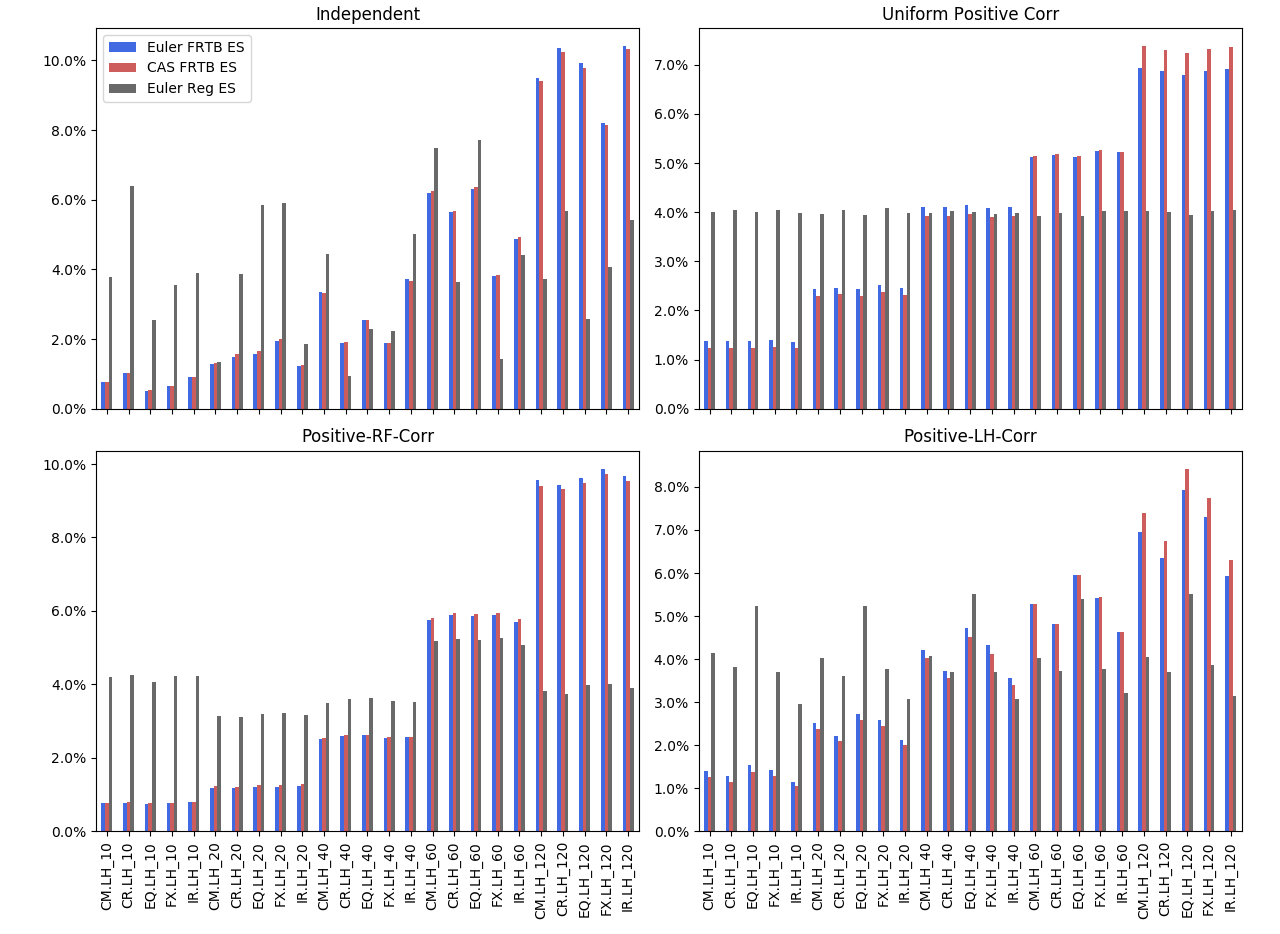}
\caption{Kernel smoothed allocations}
\label{WithKernelFigs}
\end{figure}

The upper-left panel of Figure \ref{NonKernelFigs} shows that the Euler allocations of regular ES present large variations and negative allocations even when there are no negative correlations. These features are due to the instability of the Euler allocation for regular ES or VaR, which has been documented in  \cite{yamai2002comparative}. The kernel smoothing technique (see \cite{ES2006}) can improve stability of the Euler allocation. Figure \ref{WithKernelFigs} presents the allocation results when the kernel smoothing technique is applied to each allocation method. Comparing Figures \ref{NonKernelFigs} and \ref{WithKernelFigs}, we can see that the kernel smoothing technique significantly improves the stability for the Euler allocation for the regular ES, but it is less effective on FRTB allocations. Negative allocations do not appear under both the FRTB allocations in this exercise, neither much difference between them is observed.

\subsection{Hedging}
In the second  simulation exercise, we analyse three scenarios of hedging relations: hedging between 2 \RF s (e.g. the hedging portfolio for CoCo bonds); hedging between 2 \LH s (e.g. hedging between 3 paris of FX rates with different LHs, say CNY-GBP, USD-GBP, CNY-USD); and hedging between two risk positions in the same bucket. To study the impact of hedging between liquidity horizon adjusted risk profiles, we view different  buckets  as different risk positions. In this way, $X_n(i,j) = \sqrt{\frac{LH_j-LH_{j-1}}{10}}\tilde{X}_n(i,j)$, and the correlations between different $\tilde{X}_n(i,j)$ are the same as the correlations between different $X_n(i,j)$. This allows us to focus on the impact of FRTB rules on allocations with hedging.

We consider the following three correlation structures:
\begin{enumerate}
\item[(i)] Strong hedging between EQ and IR: $\text{corr}(\tilde{X}(3, j), \tilde{X}(5, j)) =-0.99$ for any $j$ and zero correlation between all other pairs;
\item[(ii)] Strong hedging between $\LH_1$ and $\LH_2$: $\text{corr}(\tilde{X}(i,1),\tilde{X}(i,2)) =-0.99$ for all $i$ and zero correlation between all other pairs;
\item[(iii)] Strong hedging between 2 risk positions within the same bucket: $\text{corr}(\tilde{X}_1(i,j), \tilde{X}_2(i,j)) =-0.99$ for all $i,j$, and zero correlation between all other pairs.
\end{enumerate}
Intuition obtained in these three cases remains to be true when correlations are less extreme.

The simulation settings remain the same as in the previous exercise. The IMCC and regular ES are reported in Table \ref{mytable_frtb_es_test2} below. We can see from Table \ref{mytable_frtb_es_test2} that the IMCC is between 2.5 to 3.6 times to the regular ES. On the other hand, because FRTB restricts hedging among different buckets, the ratios between IMCC and ES in scenario (i) and (ii) are much larger than the ratio in scenario (iii), where hedging within the same bucket is not restricted by FRTB.

\begin{table}[h]
\centering
  \begin{tabular}{ c | c  c }
    \hline
     Scenario& IMCC & Regular ES \\ \hline
    (i) RF Hedging & 7.90 & 2.17 \\
    (ii) LH Hedging   & 8.43 &  2.55 \\
    (iii) Position Hedging & 0.84 & 0.33  \\
    \hline
    \hline
  \end{tabular}
  \caption { FRTB IMCC v.s. Regular ES}\label{mytable_frtb_es_test2}
\end{table}

\begin{figure}[!htb]
    \centering
    \begin{minipage}{.33\textwidth}
        \centering
        \includegraphics[width=1\linewidth, height=0.25\textheight]{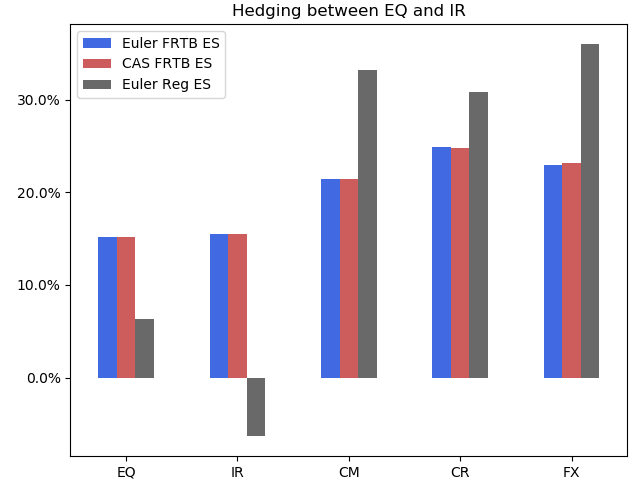}

    \end{minipage}%
    \begin{minipage}{0.33\textwidth}
        \centering
        \includegraphics[width=1\linewidth, height=0.25\textheight]{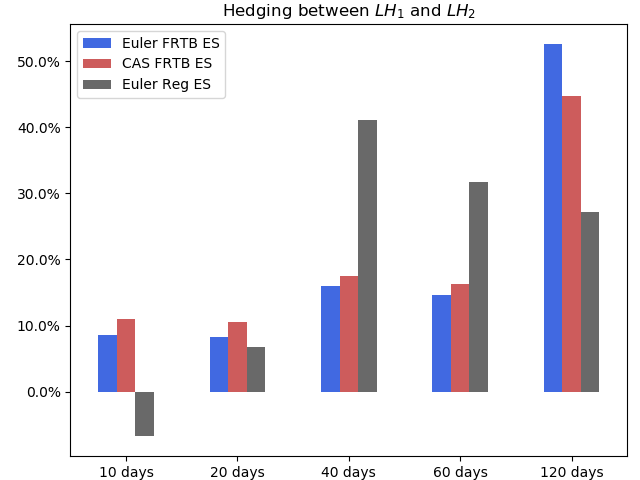}

    \end{minipage}
        \begin{minipage}{.33\textwidth}
        \centering
        \includegraphics[width=1\linewidth, height=0.25\textheight]{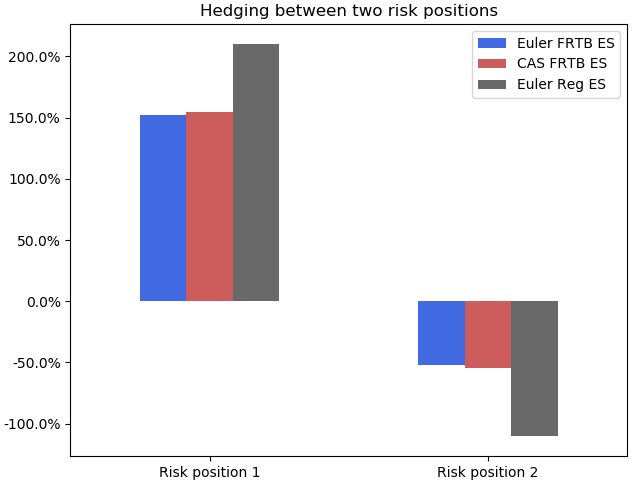}

    \end{minipage}%
    \caption{Allocations of IMCC and regular ES for portfolios with hedging components. Left panel: hedging structure (i); Middle panel: hedging structure (ii); Right panel: hedging structure (iii). Each panel presents the percentage of allocation to different $\tilde{X}(i,j)$. The total capital charges are reported in Table \ref{mytable_frtb_es_test2}.}
    \label{DeskHedging}
\end{figure}

Figure \ref{DeskHedging} illustrates different allocations of IMCC and regular ES.
The left and middle panels show that, even though there are negative correlations between different risk factor or liquidity horizon buckets, the Euler and CAS allocations of IMCC are all positive. This  confirms our analysis in Example \ref{exa:neg-all}.

When hedging appears in the same bucket, the right panel in Figure \ref{DeskHedging} shows that there could be negative allocations for both Euler and CAS allocations of IMCC. But their magnitudes are smaller than the Euler allocations of the regular ES. In the Euler allocation of the regular ES, one scenario extraction is applied to each loss simulation of  $250$ days. However, in both Euler and CAS allocations of IMCC, one scenario extraction is applied to each bucket. Therefore, there are in total $30=6\times 5$ scenario extractions applied to each loss simulation of $250$ days. Then the final allocation of a risk position is a weighted sum of $30$ scenario extraction results. Hence the FRTB allocations produce much more stable results comparing to regular ES allocations.

\begin{figure}[h]
\centering
   \includegraphics[height=0.45\textwidth,width=0.65\textwidth]{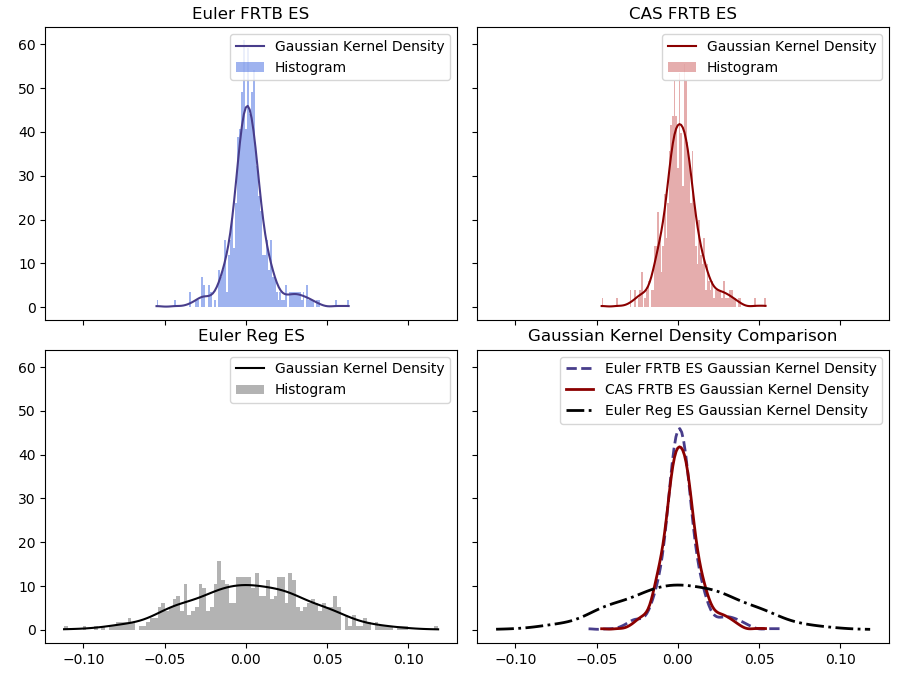}
\caption{Histograms and kernel densities for FRTB allocations and regular ES allocation. Extreme allocations: i) Euler FRTB ES: left end, -5.50\%; right end: 6.32\%; ii) CAS FRTB ES: left end, -4.69\%; right end: 5.39\%; iii) Euler Regular ES: left end, -11.19\%; right end: 11.83\%.   }
\label{DeskHistogram}
\end{figure}

In order to further analyse negativity and stability of different allocations, we extend the hedging scenario (iii) from 2 risk positions to $20$ risk positions, with each pair of risk positions following the hedging scenario (iii). We apply different allocation methods to allocate capital to each risk position and each bucket. Figure \ref{DeskHistogram} illustrates histograms and kernel densities of these allocations for each allocation method. Even without aggregation from  different risk factor and liquidity horizon classes, Figure \ref{DeskHistogram}  shows that the Euler and CAS allocations of IMCC still produce tighter histograms comparing to the case for the Euler allocation of the regular ES. Comparing the Euler and CAS allocations, we observe that the CAS allocation produces slightly more stable results with less extreme allocations. This is due to the fact that the CAS allocation is an averages of $5!$ permutations (see Definition \ref{def:CAS-IMCC}) which further improve the stability of allocations. However, the CAS FRTB allocation requires $5!$ times more computation than the Euler FRTB allocation. 


\subsection{Allocations with scaling adjustment}
In the third simulation exercise, we illustrate the impact of the choice of reduced sets on the IMCC allocations with scaling adjustment introduced in Section \ref{sec:scal-adj}. Consider the situation where the reduced factor set is chosen so that all $X_n(i,j)$ have similar distributions in the stressed period and the current period, then $\ES^{\RS}(X(i))$ is similar to $\ES^{\RC}(X(i))$, and the allocations $\ES^{\RS}\big(X_n(i,j)\,|\, X(i)\big)$ and $\ES^{\RC}\big(X_n(i,j)\,|\, X(i)\big)$ are similar as well. Therefore, the second and the third terms on the right-hand side of \eqref{IMCC-ES} are similar, so
\begin{equation}\label{IMCC-app}
 \IMCC^{\text{E,S}}\big(X_n(i,j)\,|\, X(i)\big) \approx 0.5\frac{\ES^{\RS}\big(X(i)\big)}{\ES^{\RC}\big(X(i)\big)} \ES^{\FC}\big(X_n(i,j)\,|\, X(i)\big).
\end{equation}
This allocation will be significantly different from the case where risk factors have distinct distributions in the stress period and the current period.

We follow the convention of the previous exercise where different buckets are treated as different risk positions. We consider a portfolio with two risk positions. During the current period, all $\tilde{X}_n(i,j)$ are independent and have the same distribution.  During the stress period, the correlations between any pairs of $\tilde{X}_n(i,j)$ become $0.7$. The standard deviations of $\tilde{X}_1(3,3)$ and $\tilde{X}_2(1, 4)$ during the stress period become $9$ times of the standard deviations during the current period.
Distributions of all other $\tilde{X}_n(i,j)$ in the stress period are assumed to be the same as in the current period.

We consider two reduced sets:
\begin{enumerate}
\item[Set A:] All risk factors except $60$-days EQ  and $120$-days CM;
\item[Set B:] All risk factors except $40$-days EQ and $60$-days CM.
\end{enumerate}
The reduced set B excludes risk factors which have more risky distributions in the stress period.  But the reduced set A includes them. Table \ref{mytable_1} shows that both reduced sets satisfy the requirement that $\ES^{\RC}(X(i))\geq 75\% \ES^{\FC}(X(i))$ for all $i$.

\begin{table}[h]
\centering
  \begin{tabular}{ c | c  c  c  c  c c}
    \hline
     & CM & CR&EQ & FX & IR  & Unconstrained\\ \hline
    Set A & 80\% & 100\%&97\% & 100\% & 100\%  & 95\%  \\
    Set B & 97\% & 100\% &94\% & 100\% & 100\%& 98\%  \\
    \hline
    \hline
  \end{tabular}
  \caption {Ratios between ES using the reduced set and the full set. }\label{mytable_1}
\end{table}

Table \ref{allocationRatio} shows the differences of allocations with/without stress scaling adjustments. On Set A, where volatilities are different between the stress and current periods, the allocation with stress scaling adjustment generates higher risk contributions. However, on Set B where the volatilities are equal between two periods, there is no significant difference between allocations with/without stress scaling adjustments. Moreover the total IMCC is much higher using Set A than Set B. This indicates that imposing only the requirement of $\ES^\RC(X(i))>75\% \ES^\FC(X(i))$ leaves considerable freedom for the choice of reduced sets. And such a choice significantly impacts the IMCC and its allocations. The allocation with stress scaling adjustment could effectively allocate more capital to risk factors with more risky distributions during stress periods.

\begin{table}[h]
\centering
  \begin{tabular}{ c | c  c  c  c}
    \hline
{} &     Set A &    Set A &  Set B&  Set B \\
{} & (Adjustment) & (Without adj)& (Adjustment) & (Without adj)\\
\hline
CM.60 days.Position 2 &  4.00\% &  2.24\% &          1.43\% &  1.43\% \\
EQ.40 days.Position 1 &  5.04\% &  3.26\% &          2.11\% &  2.11\% \\
\hline
\hline
\end{tabular}
  \caption {Percentages of allocations with and without stress-scaling adjustment using different reduced factor sets.  Columns labeled adjustment report allocations using \eqref{IMCC-ES}, columns labeled without adj report allocation using \eqref{Euler-IMCC}. The total IMCC are the same in both methods: IMCC(Set A)=11.55; IMCC(Set B)=3.14.}\label{allocationRatio}
\end{table}

\section{Conclusion}
We formulate the IMCC for the FRTB IMA in a mathematical framework, incorporating risk factor and liquidity horizon bucketing, liquidity horizon adjustment, and stress period scaling. We introduce two computationally efficient allocation methods for the FRTB IMCC. Simulation shows that both methods allocates more capital to risk factors with longer liquidity horizon, and produce more stable and less negative allocations than allocations under the current regulation framework. We also find that the IMCC and its allocations are sensitive to the choice of reduced set of risk factors for the stress period scaling.

\vspace{2mm}

\noindent {\bf Acknowledgements} The authors thank Udit Mahajan and Diane Pham for helpful discussions at the early stage of this project. The authors are grateful for Paul Embrechts, Demetris Lappas,  Dirk Tasche, and Ruodu Wang for their comments on the paper. The first author is grateful for Citibank for the financial support of PhD studentship at LSE.




\appendix
\section{Proofs}
\subsection{Proof of Lemma \ref{lem:IMCC-pro}}
The expected shortfall is positive homogeneous, then $\ES(a X(i,j)) = a \ES(X(i,j))$. All operations in \eqref{ES_i}, \eqref{ES-IMA}, and \eqref{IMCC} are positive homogeneous. Hence the statement in (i) holds.

For (ii), recall that the expected shortfall is sub-additive, i.e.,
\[
 \ES((X+Y)(i,j)) \leq \ES(X(i,j)) + \ES(Y(i,j)).
\]
When $\ES((X+Y)(i,j))\geq 0$ for all $j$, then
\begin{multline*}
 \ES((X+Y)(i)) = \sqrt{\sum_{j=1}^5 \ES((X+Y)(i,j))^2} \leq \sqrt{\sum_{j=1}^5 \big[\ES(X(i,j)) + \ES(Y(i,j))\big]^2}\\ \leq \sqrt{\sum_{j=1}^5 \ES(X(i,j))^2} + \sqrt{\sum_{j=1}^5 \ES(Y(i,j))^2} = \ES(X(i)) + \ES(Y(i)),
\end{multline*}
where the second inequality follows from the Minkowski inequality.

For (iii), it follows from the sub-additivity for $\ES^{\FC}$ that
\[
 \ES^{\FC}((X+Y)(i)) \leq \ES^{\FC}(X(i)) + \ES^{\FC}(Y(i)).
\]
Then, when \eqref{cond1} is satisfied, we have
\begin{align*}
 \IMCC((X+Y)(i)) = &\ES^{\FC}((X+Y)(i)) \frac{\ES^{\RS}((X+Y)(i))}{\ES^{\RC}((X+Y)(i))} \\\leq& \frac{\ES^{\RS}((X+Y)(i))}{\ES^{\RC}((X+Y)(i))} \Big[\ES^{\FC}(X(i)) + \ES^{\FC}(Y(i))\Big] \\\leq& \frac{\ES^{\RS}(X(i))}{\ES^{\RC}(X(i))} \ES^{\FC}(X(i)) + \frac{\ES^{\RS}(Y(i))}{\ES^{\RC}(Y(i))} \ES^{\FC}(Y(i))
 \\=& \IMCC(X(i)) + \IMCC(Y(i)).
\end{align*}
\qed

\subsection{Proof of Lemma \ref{lem:Euler}}
Consider fixed $n,\ i,\ j$ in Lemma \ref{lem:Euler}. Recall definitions of the FRTB ES in \eqref{ES_i} and $X^{v,j}(i)$ in \eqref{Xvj}. Using the chain rule to take the derivative with repsect to $v_n$ as in \eqref{Euler-der-0}, we obtain
\begin{align*}
\frac{\partial}{\partial v_n} \ES(X^{v,j}(i))= &\frac{\partial}{\partial v_n} \sqrt{\sum_{k\neq j} \ES(X(i,k))^2+\ES\big[\sum_{m\neq n}v_m X_m(i,j)+v_n X_n(i,j)\big]^2}\\
= &\frac{2 \ES\big[\sum_{m\neq n}v_mX_m(i,j)+v_nX_n(i,j)\big]\frac{\partial}{\partial v_n }\ES\big[\sum_{m\neq n}v_mX_m(i,j)+v_nX_n(i,j)\big]}{2\sqrt{\sum_{k\neq j} \ES(X(i,k))^2+\ES\big[\sum_{m\neq n}v_mX_m(i,j)+v_nX_n(i,j)\big]^2}} \\
= &\frac{ ES(X^v(i,j))}{ES(X^{v,j}(i))}\frac{\partial}{\partial v_n }ES(X^v(i,j)), \\
\end{align*}
where $X^v(i,j)$ is defined as in Lemma \ref{lem:Euler}. Then the proof is concluded by assigning all $v_n=1$.
\qed

\subsection{Proof of Proposition \ref{prop:Euler-IMCC}}
Since the FRTB ES, defined in \eqref{ES_i}, is a risk measure homogeneous of degree $1$. It then follows from Euler's theorem on homogeneous functions (see \cite[Theorem A.1]{tasche2007capital}) that the Euler allocation on FRTB ES is a full allocation, i.e.,
\begin{equation*}\label{Euler-ES-F}
 \sum_{n,j} \ES^{\FC}\big(X_n(i,j) \,|\, X(i)\big) = \ES^{\FC}(X(i)).
\end{equation*}
This identity, combined with \eqref{ES-IMA} and \eqref{IMCC}, yields
\begin{align*}
 \sum_{n,i,j} \IMCC\big(X_n(i,j) \,|\, X(i)\big) = &0.5 \sum_{i=1}^6 \frac{\ES^{\RS}(X(i))}{\ES^{\RC}(X(i))} \Big(\sum_{n,j} \ES^{\FC} \big(X_n(i,j) \, |\, X(i)\big)\Big) \\
 =& 0.5 \sum_{i=1}^6 \frac{\ES^{\RS}(X(i))}{\ES^{\RC}(X(i))}  \ES^{\FC}(X(i)) = \IMCC(X).
\end{align*}
\qed

\subsection{Proof of Lemma \ref{lem:CAS}}
When $\LH_j$ is in the first column of the permutation $r$, i.e., $\mathcal{L}^{-1} (r, j)=1$, the row $X^{v,r,j}(i)$ has only one nonzero entry $\sum_n v_n X_n(i,j)$ at the $j$-th column. Then
\[
 \ES\big(X^{v,r,j}(i)\big) = \big|\ES\big(\sum_n v_n X_n(i,j)\big)\big|.
\]
Since the expected shortfall is homogeneous of degree $1$, then
\begin{align*}
\partial_{v_n} \ES\big(X^{v,r,j}(i)\big)\Big|_{v=q} = &\text{sgn}\big(\ES(q X(i,j)) \big) \,\partial_{v_n} \ES\big(\sum_n v_n X_n(i,j)\big) \Big|_{v=q} \\
=& \text{sgn}\big(\ES(X(i,j)) \big) \,\partial_{v_n} \ES\big(\sum_n v_n X_n(i,j)\big) \Big|_{v=1}.
\end{align*}
As a result,
\begin{align*}
 \CAS(r, X_n(i,j)) = \int_0^1 \partial_{v_n} \ES\big(X^{v,r,j}(i)\big)\big|_{v= q} \, dq &= \int_0^1 \partial_{v_n} \ES\big(X^v(i,j)\big)\big|_{v=1} \,dq \\
 &= \partial_{v_n} \ES\big(X^v(i,j)\big)\big|_{v=1}.
\end{align*}
Note that $\eta(r, i,j) = \text{sgn}\big(\ES(q X(i,j)) \big) $ in this case. Therefore the previous expression of $\CAS(r, X_n(i,j))$ agrees with \eqref{CAS-der}.

When $\LH_j$ is not in the first column, i.e., $\mathcal{L}^{-1}(r, j)>1$,
\[
 \ES\big(X^{v, r, j}(i)\big) =\sqrt{\ES\big(\sum_n v_n X_n(i,j)\big)^2 + \sum_{1\leq s< \mathcal{L}^{-1}(r,j)} \ES\big(X(i, \mathcal{L}(r,s))\big)^2}.
\]
Denote
\[
 \ES\big(X^{q, r, j}(i)\big) =\sqrt{\ES\big(q X(i,j)\big)^2 + \sum_{1\leq s< \mathcal{L}^{-1}(r,j)} \ES\big(X(i, \mathcal{L}(r,s))\big)^2}.
\]
It follows from the homogeneous property of the expected shortfall that
\begin{align*}
\partial_{v_n} \ES\big(X^{v, r,j}(i)\big)\Big|_{v=q} = & \frac{\ES\big(q X(i,j)\big) \, \partial_{v_n} \ES\big(\sum_n v_n X_n(i,j)\big)\big|_{v=q}}{\ES\big(X^{q, r, j}(i)\big)} \\
=& \frac{q\ES\big(X(i,j)\big) \, \partial_{v_n} \ES\big(\sum_n v_n X_n(i,j)\big)\big|_{v=1}}{\ES\big(X^{q, r, j}(i)\big)}.
\end{align*}
Integrating the derivative with respect to $q$, we obtain
\begin{align*}
& \int_0^1 \partial_{v_n} \ES\big(X^{v, r,j}(i)\big)\Big|_{v=q} \, dq =\partial_{v_n} \ES\big(X^v(i,j)\big)\big|_{v=1} \int_0^1 \frac{q \ES\big(X(i,j)\big)}{\ES\big(X^{q,r,j}(i)\big)} dq\\
 &= \frac{\partial_{v_n} \ES\big(X^{v, r,j}(i)\big)\Big|_{v=1}}{\ES\big(X(i,j)\big)} \int_0^1 \frac{q \ES\big(X(i,j)\big)^2}{\ES\big(X^{q,r,j}(i)\big)} dq =  \frac{\partial_{v_n} \ES\big(X^{v, r,j}(i)\big)\Big|_{v=1}}{2\,\ES\big(X(i,j)\big)} \int_0^1 \frac{d \big(q^2 \ES\big(X(i,j)\big)^2\big)}{\ES\big(X^{q,r,j}(i)\big)}  dq\\
 & = \eta(r,i,j) \,\partial_{v_n} \ES\big(X^{v, r,j}(i)\big)\Big|_{v=1}.
\end{align*}
\qed

\subsection{Proof of Proposition \ref{prop:CAS-IMCC}}
From Lemma \ref{lem:CAS} and the fact that the standard Euler allocation is a full allocation,  we have
\begin{align*}
 &\sum_n \CAS(r, X_n(i,j)) = \eta(r,i,j) \sum_n \partial_{v_n} \ES\big(X^v(i,j)\big) \big|_{v=1}= \eta(r, i, j) \ES\big(X(i,j)\big) \\
 &= \sqrt{\sum_{1\leq s\leq \mathcal{L}^{-1}(r, j)} \ES\big(X(i, \mathcal{L}(r,s))\big)^2} - \sqrt{\sum_{1\leq s < \mathcal{L}^{-1} (r, j)} \ES\big(X(i, \mathcal{L}(r, s))\big)^2}.
\end{align*}
Therefore
\[
 \sum_{n,j} \CAS(r, X_n(i,j)) = \ES(X(i)).
\]
The rest proof is similar to the proof of Proposition \ref{prop:Euler-IMCC}. \qed

\subsection{Proof of Proposition \ref{prop:stressallocationnewlyadded}}
Recall $X^{v,j}(i)$ in \eqref{Xvj}. The right-hand side of Definition \ref{den:defn3-11} then yields
\begin{align*}
0.5 \frac{\partial}{\partial v_n} \Big[&\frac{\ES^{\RS} \big(X^{v,j}(i)\big)}{\ES^{\RC} \big(X^{v,j}(i)\big)} \ES^{\FC}\big(X^{v,j}(i)\big)\Big] \Big|_{v=1}\\
= 0.5 \Big[&\frac{\ES^{\RS}(X^{v,j}(i))}{\ES^{\RC}(X^{v,j}(i))} \, \frac{\partial}{\partial v_n} \ES^{\FC} (X^{v,j}(i))\\
+ & \frac{\ES^{\FC}(X^{v,j}(i))}{\ES^{\RC}(X^{v,j}(i))} \, \frac{\partial}{\partial v_n} \ES^{\RS} (X^{v,j}(i)) \\
- &\frac{\ES^{\RS}(X^{v,j}(i))\ES^{\FC}(X^{v,j}(i))}{\ES^{\RC}(X^{v,j}(i))^2} \, \frac{\partial}{\partial v_n} \ES^{\RC} (X^{v,j}(i))\Big]\Big|_{v=1}.
\end{align*}
Noticing that
\[
\ES(X^{v,j}(i))\big|_{v=1}=\ES(X(i)),
\]
and from \eqref{Euler-der-0} that
\[
 \frac{\partial}{\partial v_n} \ES (X^{v,j}(i))\big|_{v=1}=\ES(X_n(i,j)|X(i));
\]
substituting these identities into the equation above, we verify the statement.
\qed

\subsection{Proof of Proposition \ref{prop:Euler-CAS-ad}}
Recall that
\[
 \sum_{n,j} \ES\big(X_n(i,j) \,|\, X(i)\big) = \ES\big(X(i)\big).
\]
Then applying the previous identity to the Euler allocation for $\ES^{\FC}, \ES^{\RS}$, and $\ES^{\RC}$, respectively, we obtain
\begin{align*}
\sum_{n,j} \IMCC^{\text{E,S}}\big(X_n(i,j) \,|\, X(i)\big) = 0.5 \Big[&\frac{\ES^{\RS}(X(i))}{\ES^{\RC}(X(i))} \, \ES^{\FC} \big(X(i)\big) + \frac{\ES^{\FC}(X(i))}{\ES^{\RC}(X(i))} \, \ES^{\RS} \big(X(i)\big) \\-& \frac{\ES^{\RS}(X(i))\ES^{\FC}(X(i))}{\ES^{\RC}(X(i))^2} \, \ES^{\RC} \big(X(i)\big)\Big]\\
= 0.5\,& \frac{\ES^{\RS}(X(i))}{\ES^{\RC}(X(i))} \, \ES^{\FC} \big(X(i)\big)= \IMCC\big(X(i)\big).
\end{align*}
The proof for $\IMCC^{\text{C,S}}$ is similar. \qed

\bibliographystyle{plain}
\bibliography{reff}

\begin{thebibliography}{10}

\bibitem{aumann1974values}
Robert~J Aumann and Lloyd~S Shapley.
\newblock {\em Values of non-atomic games}.
\newblock Princeton University Press, 1974.

\bibitem{denault2001coherent}
Michel Denault.
\newblock Coherent allocation of risk capital.
\newblock {\em Journal of risk}, 4:1--34, 2001.

\bibitem{ES2006}
Eduardo Epperlein and Alan Smillie.
\newblock Portfolio risk analysis cracking var with kernels.
\newblock {\em Risk Magazine}, 19(8):70, 2006.

\bibitem{li2016organising}
Yadong Li, Marco Naldi, Jeffrey Nisen, and Yixi Shi.
\newblock Organising the allocation.
\newblock {\em Risk Magazine}, 2016.

\bibitem{litterman1996hot}
Robert Litterman.
\newblock Hot spots? and hedges.
\newblock {\em The Journal of Portfolio Management}, 23(5):52--75, 1996.

\bibitem{Basel25}
Basel~Committee on~Banking~Supervision.
\newblock {\em Revisions to the {B}asel II {M}arket {R}isk {F}ramework}.
\newblock BCBS, 2009.

\bibitem{QIS2014}
Basel~Committee on~Banking~Supervision.
\newblock {\em Fundamental {R}eview of the {T}rading {B}ook - {I}nterim
  {I}mpact {A}nalysis}.
\newblock BCBS, 2015.

\bibitem{BCBS2016}
Basel~Committee on~Banking~Supervision.
\newblock {\em Minimum {C}apital {R}equirements for {M}arket {R}isk}.
\newblock BCBS, 2016.

\bibitem{QIS2017}
Basel~Committee on~Banking~Supervision.
\newblock {\em Basel {III} {M}onitoring {R}eport}.
\newblock BCBS, 2017.

\bibitem{BCBS2018}
Basel~Committee on~Banking~Supervision.
\newblock {\em {C}onsultative {D}ocument: Revisions to the minimum capital
  requirements for market risk}.
\newblock BCBS, 2018.

\bibitem{shapley1953}
L.~S. Shapley.
\newblock A value for {$n$}-person games.
\newblock In {\em Contributions to the theory of games, vol. 2}, Annals of
  Mathematics Studies, no. 28, pages 307--317. Princeton University Press,
  Princeton, N. J., 1953.

\bibitem{tasche1999risk}
Dirk Tasche.
\newblock Risk contributions and performance measurement.
\newblock {\em Report of the Lehrstuhl f{\"u}r mathematische Statistik, TU
  M{\"u}nchen}, 1999.

\bibitem{tasche2007capital}
Dirk Tasche.
\newblock Capital allocation to business units and sub-portfolios: the euler
  principle.
\newblock {\em Pillar II in the New Basel Accord: The Challenge of Economic
  Capital}, pages 423--453, 2008.

\bibitem{yamai2002comparative}
Yasuhiro Yamai, Toshinao Yoshiba, et~al.
\newblock Comparative analyses of expected shortfall and value-at-risk: their
  estimation error, decomposition, and optimization.
\newblock {\em Monetary and economic studies}, 20(1):87--121, 2002.

\end{thebibliography}

\end{document}